\documentclass{llncs}
\usepackage{times}
\usepackage{helvet}
\usepackage{courier}
\usepackage{xspace}
\usepackage{amssymb}
\usepackage{amsmath}
\usepackage{epic}
\usepackage{graphicx}
\usepackage{subfigure}
\usepackage[ruled,linesnumbered,boxed]{algorithm2e}
\usepackage{multirow}
\usepackage{hyperref}

\begin{document}
%
\title{Deceased Organ Matching in Australia}
\author{Toby Walsh}
\institute{UNSW Sydney, Data61, and TU Berlin\\ 
Email: tw@cse.unsw.edu.au}

\maketitle
\begin{abstract}
Despite efforts to increase the supply of organs from living
donors, most kidney transplants performed in Australia
still come from deceased donors. The age of these donated organs
has increased substantially in recent decades as the rate of fatal
accidents on roads has fallen. The Organ and Tissue Authority in
Australia
is therefore looking to design a new mechanism that
better matches the age of the organ to the age of the patient.
I discuss the design, axiomatics and performance of several
candidate mechanisms that respect the special online nature
of this fair division problem. 
\end{abstract}

\newtheorem{mydefinition}{Definition}
\newtheorem{mytheorem}{Theorem}
\newtheorem{prop}{Proposition}
\newtheorem{mylemma}{Lemma}
\newtheorem{myexample}{Example}{\bf}{\it}
\newtheorem{mytheorem1}{Theorem}
\newcommand{\myproof}{\noindent {\bf Proof:\ \ }}
\newcommand{\myqed}{\mbox{$\blacksquare$}}
\newcommand{\myend}{\mbox{$\clubsuit$}}

\newcommand{\ra}{\mbox{$\rightarrow$}}
\newcommand{\mymod}{\mbox{\rm mod}}
\newcommand{\mymin}{\mbox{\rm min}}
\newcommand{\mymax}{\mbox{\rm max}}
\newcommand{\range}{\mbox{\sc Range}}
\newcommand{\roots}{\mbox{\sc Roots}}
\newcommand{\myiff}{\mbox{\rm iff}}
\newcommand{\alldifferent}{\mbox{\sc AllDifferent}}
\newcommand{\permutation}{\mbox{\sc Permutation}}
\newcommand{\pecedence}{\mbox{\sc Perecendece}}
\newcommand{\disjoint}{\mbox{\sc Disjoint}}
\newcommand{\cardpath}{\mbox{\sc CardPath}}
\newcommand{\CARDPATH}{\mbox{\sc CardPath}}
\newcommand{\common}{\mbox{\sc Common}}
\newcommand{\uses}{\mbox{\sc Uses}}
\newcommand{\lex}{\mbox{\sc Lex}}
\newcommand{\usedby}{\mbox{\sc UsedBy}}
\newcommand{\nvalue}{\mbox{\sc NValue}}
\newcommand{\slide}{\mbox{\sc CardPath}}
\newcommand{\sliden}{\mbox{\sc AllPath}}
\newcommand{\SLIDE}{\mbox{\sc CardPath}}
\newcommand{\circularslide}{\mbox{\sc CardPath}_{\rm O}}
\newcommand{\among}{\mbox{\sc Among}}
\newcommand{\mysum}{\mbox{\sc MySum}}
\newcommand{\amongseq}{\mbox{\sc AmongSeq}}
\newcommand{\atmost}{\mbox{\sc AtMost}}
\newcommand{\atleast}{\mbox{\sc AtLeast}}
\newcommand{\element}{\mbox{\sc Element}}
\newcommand{\gcc}{\mbox{\sc Gcc}}
\newcommand{\gsc}{\mbox{\sc Gsc}}
\newcommand{\contiguity}{\mbox{\sc Contiguity}}
\newcommand{\PRECEDENCE}{\mbox{\sc Precedence}}
\newcommand{\assignnvalues}{\mbox{\sc Assign\&NValues}}
\newcommand{\linksettobooleans}{\mbox{\sc LinkSet2Booleans}}
\newcommand{\domain}{\mbox{\sc Domain}}
\newcommand{\symalldiff}{\mbox{\sc SymAllDiff}}
\newcommand{\alldiff}{\mbox{\sc AllDiff}}
\newcommand{\doublelex}{\mbox{\sc DoubleLex}}

\newcommand{\dyn}{\mbox{\sc Dyn}}
\newcommand{\preced}{\mbox{\sc Prec}}
\newcommand{\prep}{\mbox{\sc Prep}}
\newcommand{\precedsh}{\mbox{\sc $\preced_{sh}$}}
\newcommand{\precedprep}{\mbox{\sc $\preced+\prep$}}
\newcommand{\precedprepsh}{\mbox{\sc $\preced+\prep_{sh}$}}

\newcommand{\slidingsum}{\mbox{\sc SlidingSum}}
\newcommand{\MaxIndex}{\mbox{\sc MaxIndex}}
\newcommand{\REGULAR}{\mbox{\sc Regular}}
\newcommand{\regular}{\mbox{\sc Regular}}
\newcommand{\precedence}{\mbox{\sc Precedence}}
\newcommand{\STRETCH}{\mbox{\sc Stretch}}
\newcommand{\SLIDEOR}{\mbox{\sc SlideOr}}
\newcommand{\NAE}{\mbox{\sc NotAllEqual}}
\newcommand{\mytheta}{\mbox{$\theta_1$}}
\newcommand{\mysigma}{\mbox{$\sigma_2$}}
\newcommand{\myoverline}[1]{\mbox{$\bar{#1}$}}

\newcommand{\mysigmatwo}{\mbox{$\sigma_1$}}
\newcommand{\todo}[1]{{\tt (... #1 ...)}}
\newcommand{\myOmit}[1]{}
\newcommand{\dpsb}{DPSB}

\newcommand{\mm}{{\mathcal M}}
\newcommand{\ma}{{\mathcal A}}
\newcommand{\mi}{{\mathcal I}}

\section{Introduction}

Kidney disease is a major problem in Australia. 
Thousands of people are on dialysis. Many spend years
waiting for a transplant, each costing the 
health care budget hundreds of thousands
of dollars. In addition,
as dialysis takes up several days each week,
many are unable to 
work and depend on support from the state. 
The total cost to the Australian economy 
runs into billions of dollars annually. 
In 2016, 85\% of transplants involved
a kidney coming from a deceased person,
whilst only 15\% of transplants came
from a living donor. 
Whilst there has been considerable focus
in the literature of late on increasing the supply of organs
by developing mechanisms for paired exachange,
only 2.5\% of these living
donations came from paired exchange. 
Most living donors were a spouse,
family member or friend of the recipient. 

Organs coming from deceased people still provide
the majority of all transplanted kidneys. 
Many come from people
killed in road traffic accidents. 
Matching such organs to patients on the waiting list 
is becoming more challenging
as roads become safer. 
The mean age of 
donated organs has increased from 32 years in 1989 to 46 years in 2014.
Advances in medicine mean that doctors are 
also now willing to transplant older kidneys. In 2014,
the oldest organ transplanted came from a person
who was 80 years old. This compares to 1989, the first
year for which records are available, when the
oldest organ transplated came from a person aged just 69. 
The Organ and Tissue Authority of Australia,
the government body charged with the task
of allocating organs to patients fairly and efficiently,
is therefore looking to develop a new matching 
mechanism. Their goal is to develop a new procedure which 
matches the
age of the patients and organs.

\section{Organ matching mechanisms}

The mechanism used at present in Australia does not explicitly
take age of the patients or organs into account. As a result, 
young organs will be offered to old patients,
and old organs to young patients. 
Neither are very desirable. 
Even if an old patient would like a 
young organ, from a societal perspective,
this is not a very good outcome. The old
patient will die from natural causes with
an organ inside them that could have 
continued to function in a younger patient. 
And transplanting an old organ into a young
patient is not a good outcome for both the
individual or society. The graft will likely
fail after a few years, meaning the patient
will need a new transplant. In addition, 
the patient's immune system will now 
be highly sensitized, so that a new match
will be more difficult. 

The Organ and Tissue Authority 
is looking therefore to develop a new mechanism
in which organs are ranked by the Kidney Donor Patient Index (KDPI).
This is an integer from 0 to 100 that is
calculated from the age of the donor,
and a number of other factors like
diabetic status. A donated organ with a KDPI of X\% has an
expected risk of graft failure greater
than X\% of all donated organs. 
Similarly the Organ and Tissue Authority
wish to rank patients waiting
transplant with
the Expected Post-Transplant Survival (EPTS) score. 
This is also an integer from 0 to 100 that is
calculated from the age of the recipient,
and a number of other factors like diabetic status, and time 
on dialysis. A patient on the waiting list with a lower
EPTS is expected to have more
years of graft function from high-longevity kidneys compared to candidates with higher EPTS scores. 
Our goal is to provide the Organ
and Tissue Authority with a new mechanism
that is fair and efficient,
matching organs so that the KDPI of an arriving organ
is as close as possible to the EPTS score of their allocated
patient.

\section{Other applications}

This work fits into a broader research programme
to design mechanisms for resource
allocation problems that better reflect the complexity 
and richness of the real world \cite{wki2014,waaai2015}. 
Unlike traditional resource allocation problems \cite{rasurvey},
one of the fundamental features of 
the deceased organ matching 
problem is that it is online. We do not
know when organs will arrive to be match.
And we must match and transplant
them shortly after they arrive, before we know
what organs or patients will arrive in the
future. At the end of the year, we could
find an optimal allocation. However, we do not
have the luxury of 
waiting till the end of the year as organs
must be transplanted immediately. 
There are many other domains where 
resources are allocated in a similar online
manner. A food bank might start allocating
and distributing food to charities soon after it is donated
\cite{aagwijcai2015}.
An airport must start allocating landing slots
before all demands are known. 
A particle accelerator might
start allocating beam time 
before all requests have come
in. An university might allocate rooms
to students for the current term,
not knowing what rooms might be 
wanted in future terms. 
This work offers a case
study in how we can efficiently and fairly
solve such {\em online} allocation 
problems.
We study axiomatic properties of such
online fair division problems, 
as well as run experiments on 
real world organ data \cite{mswijcai17}.
Axiomatic analysis covers 
such properties as fairness and efficiency
(e.g. 
\cite{indivisible}-\cite{akwxecai16}), 
as well as strategic behaviour and manipulation
(e.g. 
\cite{rraamas2010}-\cite{waaai16}). 
Insights from this research may prove valuable in a range of other
domains. In future, we plan to identify and study
phase transition behaviour
\cite{hogg1}-\cite{easy-hard} which has proved
valuable in a wide range of computational domains
\cite{isai95}-\cite{waaai2002}
including social choice \cite{wijcai09,wecai10,wjair11}.

\bibliographystyle{splncs} 

\bibliography{/Users/tw/Documents/biblio/a-z,/Users/tw/Documents/biblio/a-z2,/Users/tw/Documents/biblio/pub,/Users/tw/Documents/biblio/pub2}

\end{document}